# CONSERVATION OF ANGULAR MOMENTUM AND THE EXISTENCE OF ABSOLUTE TIME AND SPACE

A. Paglietti

University of Cagliari, 09123 Cagliari, Italy

E-mail: paglietti@unica.it

**Abstract.** The law of balance of angular momentum is shown to imply the existence of absolute time, a fundamental physical quantity that is independent of the motion or position of the observer. Absolute time implies the notion of absolute simultaneity, which in turn leads to the notion of absolute distance between two points. The existence of absolute space follows as a consequence. These concepts apply to every field of physics to which the angular momentum balance law applies, and in particular to the theory of special relativity. The paper also shows that in a vacuum, the independence of the speed of light from the motion of its source makes it possible to determine the absolute positions of all points in space. The same independence also allows us to determine the state of absolute rest or motion of a reference frame from within the frame itself.

**Keywords**

Absolute time; absolute space; absolute simultaneity; angular momentum balance; Lorentz time.

# 1. Introduction

A body that spins around a barycentric axis maintains its spin angular velocity unaltered if free from torques about its centre of mass. This behaviour is a consequence of the law of conservation of angular momentum, the validity of which is beyond question here. The body's centre of mass can move through space at any speed and acceleration along any trajectory that is compatible with the applied forces and fields. The spin angular velocity of the body, as opposed to its orbital angular velocity, remains constant if the applied actions do not produce a net torque about the body's centre of mass. This behaviour implies the existence of absolute time and absolute space in every field of physics where the law of conservation of angular momentum holds true, as demonstrated in this paper. Classical physics and special relativity are two prominent examples. Because of a lack of relevant studies on the subject, similar conclusions for the fields of general relativity and quantum mechanics appear premature.

Time is the physical quantity that is measured by a clock; any properly designed clock will do. The clock we refer to here is a *spin clock*, which consists of a rotor—a circular cylindrical body of finite height—spinning at a constant angular velocity about its axis. The rotation angle of the rotor provides a measure of time that is independent of the motion of the rotor's centre of mass. This means that two initially synchronous rotors remain synchronous after being transported at different velocities through different paths in space, regardless of their relative motion. Two synchronised spin clocks can therefore provide an absolute and universal measure of the time interval between two events occurring at the points at which the clocks are placed, regardless of the time lag between the events, the distance between them, and their relative velocity.

In particular, the measurement of time from a spin clock is independent of the velocity of the inertial reference frame to which the spin clock is attached. The paper will show that the spin clock measures absolute time, as postulated in classical mechanics and involved in the Galilean transformation for inertial frames. This point is covered in detail in Section 3. The analysis is preceded by a review, presented in Section 2, of the properties of the angular momentum balance, from which the independence of the spin of a body from the motion of its centre of mass follows.

Section 4 shows that the measure of time provided by a spin clock is invariant under the Lorentz transformation (Lorentz boost). This property means that the notion of absolute time can be used within the special theory of relativity. The consequence is that relative time, as defined by the Lorentz transformation, simply becomes a particular time parameter that depends on absolute time and position. If it is used to measure speed, this parameter makes the speed of light the same in all inertial frames.

The notion of absolute simultaneity follows from the possibility of measuring absolute time at any point in space, regardless of the observer's velocity. Section 5 shows that absolute simultaneity implies absolute distance, and hence absolute space. In other words, absolute time implies absolute space.

Finally, the analysis in Section 6 shows that the constancy of the speed of light in a vacuum implies that the *absolute* position of a point can, at least in principle, be determined from any other point. Reference to absolute time is essential for this. The same section also shows that the absolute motion of an inertial frame is theoretically detectable by an observer inside the frame, without requiring any external point as reference. This result is a consequence of the fact that in a vacuum, the velocity of light is independent of the velocity of its source, as predicted by Maxwell's equations and in agreement with the experimental evidence.

The relationship between the angular momentum balance law and absolute time does not seem to have been considered previously, and for this reason, a review of the literature on this argument cannot be provided. All necessary references are cited in the text where appropriate.

## 2. Preliminaries from classical mechanics

Let $\mathcal{B}$ denote a material body in ordinary Euclidean space, and let $O$ be a point in this space. To simplify the formulae, we take $O$ as the origin of the reference frame adopted here to describe the position of $\mathcal{B}$. The angular momentum of $\mathcal{B}$ with respect to $O$ is therefore expressed by the classical formula

$$\boldsymbol{L}_O = \int_{\mathcal{B}} \boldsymbol{r} \times \dot{\boldsymbol{r}} \, dm, \tag{1}$$

where $\boldsymbol{r}$ is the position vector of the mass element $dm$ and $\dot{\boldsymbol{r}}$ the velocity vector of the same element in the considered reference frame (Fig. 1). The capital letter appended to vector $\boldsymbol{L}$ indicates the point about which the angular momentum is taken.

If $O$, and thus the attached reference frame, move, the angular momentum $\boldsymbol{L}_O$ changes because $\boldsymbol{r}$ and $\dot{\boldsymbol{r}}$ change. An important result of classical mechanics states that the angular momentum of a body with respect to a point, say $O$, can be decomposed into two angular momenta with respect to $O$ and $C$, respectively, according to the relation:

$$\boldsymbol{L}_O = \boldsymbol{L}_O^{orb} + \boldsymbol{L}_C^{spin}, \tag{2}$$

where $C$ is the centre of mass of $\mathcal{B}$ (Fig.1). The component $\boldsymbol{L}_O^{orb}$ appearing in this equation is the *orbital* angular momentum of $\mathcal{B}$ with respect to $O$. It is given by

$$\boldsymbol{L}_O^{orb} = \boldsymbol{R} \times m \dot{\boldsymbol{R}}, \tag{3}$$

where $m$ denotes the total mass of $\mathcal{B}$, while $\boldsymbol{R}$ is the position vector of $C$ with respect to $O$. It represents the angular momentum about $O$ of the total mass of $\mathcal{B}$, which is assumed to be concentrated at the centre of mass $C$ and moving with it. $\boldsymbol{L}_O^{orb}$ depends on the frame of reference and its motion, since so do $\boldsymbol{R}$ and $\dot{\boldsymbol{R}}$.

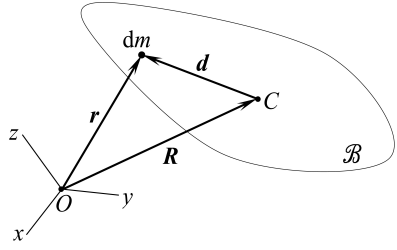

**FIG. 1.** Centre of mass $C$ of a body $\mathcal{B}$ and position vectors $\boldsymbol{r}$ and $\boldsymbol{d}$ of a mass element d$m$, relative to $O$ and $C$, respectively. The position vector of $C$ relative to $O$ is denoted by $\boldsymbol{R}$.

The other component of $\boldsymbol{L}_O$ that appears in Eq. (2) is $\boldsymbol{L}_C^{spin}$. This is referred to as the *intrinsic* angular momentum, or *spin* angular momentum (or simply *spin*), and is defined as

$$\boldsymbol{L}_C^{spin} = \int_{\mathcal{B}} \boldsymbol{d} \times \dot{\boldsymbol{r}}\, \mathrm{d}m = \int_{\mathcal{B}} \boldsymbol{d} \times \dot{\boldsymbol{d}}\, \mathrm{d}m, \qquad (4)$$

where

$$\boldsymbol{d} = \boldsymbol{r} - \boldsymbol{R} \qquad (5)$$

is the position vector of d$m$ relative to the body's centre of mass $C$ (cf. Fig. 1). The second equality in Eq. (4) follows from the fact that

$$\dot{\boldsymbol{r}} = \dot{\boldsymbol{R}} + \dot{\boldsymbol{d}}, \qquad (6)$$

and that, moreover,

$$\int_{\mathcal{B}} \boldsymbol{d}\, \mathrm{d}m = 0, \qquad (7)$$

since point $C$ is the centre of mass of $\mathcal{B}$ (for more details on this point cf., e.g., p. 9 of [1]).

Since Eq. (4)$_2$ only involves relative distances from the centre of mass, $\boldsymbol{L}_C^{spin}$ does not depend on the motion or position of $C$ in space. This fact makes $\boldsymbol{L}_C^{spin}$ invariant relative to translations of the reference frame and insensitive to the velocities and accelerations of $C$, which justifies the term *intrinsic* attributed to this component of $\boldsymbol{L}_O$.

The angular momentum balance law (Euler's second law of motion) states that:

$$\dot{\boldsymbol{L}}_O = \frac{\mathrm{d}\boldsymbol{L}_O}{\mathrm{d}t} = \boldsymbol{\Gamma}_O, \qquad (8)$$

where $\boldsymbol{\Gamma}_O$ denotes the total torque about point $O$ resulting from all the forces and torques acting on $\mathcal{B}$.

In particular, $O$ can be assumed to coincide with the centre of mass $C$ of the body. This causes the orbital component of the angular momentum to vanish, as follows immediately from Eq. (3), since $\boldsymbol{R} = \boldsymbol{0}$ in this case. From Eqs. (1) and (8), we therefore conclude that:

$$\dot{\boldsymbol{L}}_C^{spin} = \frac{d\boldsymbol{L}_C^{spin}}{dt} = \boldsymbol{\Gamma}_C \,. \tag{9}$$

Here, $\boldsymbol{\Gamma}_C$ is the value of $\boldsymbol{\Gamma}_O$ relative to point $C$.

The validity of Eq. (9) is general. It states that a change in $\boldsymbol{L}_C^{spin}$ can only be produced by applying to $\mathcal{B}$ a torque or a system of forces that results in a non-vanishing torque about the centre of mass of the body. From Eq. (9), it follows that if $\boldsymbol{\Gamma}_C$ vanishes (i.e., if no torque is applied to $\mathcal{B}$), then the intrinsic spin remains constant, regardless of the motion of the body and its centre of mass.

Expressed in component form, Eq. (9) yields

$$\dot{L}_{C\,x}^{spin} = \Gamma_{C\,x}\,,$$
$$\dot{L}_{C\,y}^{spin} = \Gamma_{C\,y}\,, \tag{10}$$
$$\dot{L}_{C\,z}^{spin} = \Gamma_{C\,z}\,.$$

These equations show that a component of $\boldsymbol{L}_C^{spin}$ along a given coordinate axis can only be modified by a component of $\boldsymbol{\Gamma}_C$ along the same axis. In particular, this means that a torque normal to the axis of spin of a spinning body leaves the magnitude of the angular momentum about the same axis unaltered.

A full discussion of the topic addressed in this section, as applied to a system of particles, can be found in Refs. [2] to [5]. The case of continuous bodies follows directly from that of a system of particles, and is treated explicitly in [6].

## 3. Absolute time and absolute simultaneity

In classical mechanics, time is assumed to be universal and absolute: it runs at the same rate everywhere in the universe, regardless of the observer's motion or position. The spin clock introduced in Section 1 can be used to prove the existence of absolute time and to measure absolute time intervals. The proof stems from the fact that the spin angular momentum of the rotor of a spin clock does not change if the rotor is kept free from torques acting in the direction of its axis. As shown in the previous section, this result follows from Eq. (9), and is ultimately the consequence of the angular momentum balance law.

Quite generally, the spin angular momentum of the rotor of a spin clock can be expressed as

$$\boldsymbol{L}_C^{spin} = I\,\boldsymbol{\omega}\,, \tag{11}$$

where $\boldsymbol{\omega}$ is the angular velocity vector of the rotor about its axis, while $I$ is its moment of inertia about the same axis. For a solid cylindrical rotor of radius $r_{rot}$ and mass $M$, we have that

$$I = \frac{1}{2} M \, r_{rot}^{\,2} \,. \tag{12}$$

In particular, in a reference system with the origin at the centre of mass of the rotor and the $x$-axis coinciding with the rotor's axis, the components of $\boldsymbol{\omega}$ are:

$$\omega_x = \dot{\theta} \quad \text{and} \quad \omega_y = \omega_z = 0 \,, \tag{13}$$

where $\theta$ denotes the rotation angle of the rotor about its axis. In that reference system, Eq. (11) yields

$$L^{spin}_{C\,x} = I\,\dot{\theta} \quad \text{and} \quad L^{spin}_{C\,y} = L^{spin}_{C\,z} = 0 \,. \tag{14}$$

By introducing Eqs (14)$_1$ into Eqs (10)$_1$, we infer that $\ddot{\theta} = 0$ if the $x$-component of the torque acting on the rotor vanishes (i.e., if $\Gamma_{C\,x} = 0$). This means that, in the present case,

$$\dot{\theta} = \omega_{clk} = const \,. \tag{15}$$

The constant $\omega_{clk}$ introduced in this equation is a machine constant. It represents the rotor's spin angular speed at which the clock works once isolated from torque components along the rotor's axis. In theory, isolating the rotor is as simple as mounting its shaft on frictionless bearings. In real life, the clock must also incorporate a device to compensate for the inevitable friction losses. No matter how we move a spin clock through space, its rotor will spin at the same angular speed $\omega_{clk}$ as long as no net torque is applied to it in the direction of its axis, in agreement with what is stated by Eq. (15).

From the same Eq (15), we obtain

$$\mathrm{d}\theta = \omega_{clk}\,\mathrm{d}t \,, \tag{16}$$

which demonstrates that $\theta$ can be used as a consistent measure of time since the change in $\theta$ is proportional to the change in $t$. The time under consideration is absolute time because Eq. (16) is derived from the angular momentum balance law, which refers to absolute time. The law of angular momentum balance implies, therefore, the possibility of constructing an absolute time clock. The time thus measured is defined to within an inessential additive constant that represents the value of $\theta$ at $t = 0$ and can be assigned arbitrarily.

An immediate consequence of the above analysis is that two spin clocks can be synchronised and transferred to different points in space without losing their synchronisation. The only restriction on the transfer process is that it should not involve applying a net torque component to the rotor along the direction of its axis. The destinations of the two clocks may be in relative motion, as when the clocks are transferred to different inertial frames moving at different speeds or in different directions. The two clocks can therefore measure the absolute times of events occurring at their positions in the different reference frames in which they are placed.

The absolute simultaneity of two events occurs when the clocks at the locations of the events measure the same absolute time. Absolute simultaneity is independent of the observer, because so is the time measure that the spin clocks provide. Thus, by implying the validity of Eq. (15), the angular momentum balance law implies the possibility of determining the absolute simultaneity of two events.

In the following section, it is shown that the same conclusion holds even if the observers are moving at relativistic speeds.

**4. Invariance of absolute time under the Lorentz transformation**

The angular velocity $\dot{\theta}$ of the rotor of a spin clock remains constant even if the clock is fixed to a reference frame that moves at relativistic speeds. To prove this, let us consider an observer $S$ in a reference frame $[x, y, z]$ at rest relative to the centre of mass of the rotor, and another observer $S'$ in a reference frame $[x', y', z']$ moving relative to $S$ with velocity $v$ in the positive direction of the $x$-axis. Without loss of generality, we assume that the axes $x$, $y$ and $z$ are parallel to the axes $x'$, $y'$ and $z'$, respectively. Moreover, the origins of the two reference systems are assumed to coincide for $t = t' = 0$. Here, $t$ and $t'$ are the measures of time in observer $S$ and $S'$, respectively. With these choices, the Lorentz transformation for the space-time coordinates assigned by the two observers to a given event assumes the simple form:

$$x' = \frac{x - vt}{\sqrt{1 - v^2/c^2}}, \qquad (17)$$

$$y' = y, \qquad (18)$$

$$z' = z, \qquad (19)$$

$$t' = \frac{t - vx/c^2}{\sqrt{1 - v^2/c^2}}. \qquad (20)$$

In general, at any given time, the observers $S$ and $S'$ measure different angular momenta for the same body, because the velocity of the body is different for the two observers due to their relative motion. However, a known result of the special theory of relativity states that the component of the angular momentum parallel to the direction of relative motion of the two observers—the direction of the $x$-axis in the present case—does not change under the Lorentz transformation (cf., e.g., [7], pp. 42-45, or [8], pp. 137-138). This component of the angular momentum, therefore, has the same value for both observers, irrespective of their relative velocity.

We now fix the spin clock to the frame $[x', y', z']$ so that the rotor's axis coincides with the axes $x$ and $x'$. With the clock thus positioned, consider the angular momentum of the rotor with respect to its centre of mass $C$. From Eq. (3), we infer that this angular momentum does not include any orbital component since $\boldsymbol{R} = \boldsymbol{0}$, because the angular moment is taken relative to the body's centre of mass. (Vector $\boldsymbol{R}$ represents the distance between the body's centre of mass and the point relative to which the angular momentum is being calculated, as shown in Fig. 2.) The considered angular momentum therefore coincides with the spin angular momentum $\boldsymbol{L}_C^{spin}$ of the rotor. Since the rotor's axis is parallel to the *x*-axis, the only non-vanishing component of $\boldsymbol{L}_C^{spin}$ is the one directed along the *x*-axis, which is given by:

$$L_{C\ x}^{spin} = I\,\dot{\theta}, \tag{21}$$

cf. Eq. (14). Due to the special relativity result mentioned in the preceding paragraph, this component is independent of the observers' relative velocity, $v$.

On the other hand, if $K^{spin}$ denotes the rotational kinetic energy of the rotor of the spin clock, we have that:

$$K^{spin} = \frac{1}{2} I\,\dot{\theta}^2\ . \tag{22}$$

In view of Eq. (14)$_1$, this equation can also be written as

$$K^{spin} = \frac{1}{2}\dot{\theta}\,L_{C\ x}^{spin}\ . \tag{23}$$

As observed in [9] and [10], the rotational kinetic energy of a body, *i.e.*, its kinetic energy with respect to a frame at rest relative to the body's centre of mass, is part of the internal energy of the body. As a result, the same kinetic energy is independent of the velocity of the reference frame to which the body is fixed. This means that $K^{spin}$ is independent of the velocity $v$ of the frame to which the spin clock is fixed.

From Eq. (23), it follows that

$$\dot{\theta} = \frac{2K^{spin}}{L_{C\ x}^{spin}}, \tag{24}$$

which shows that $\dot{\theta}$ must be independent of $v$, since the same applies to $K^{spin}$ and $L_{C\ x}^{spin}$. Since $\theta$ can be taken as a measure of time, we can conclude that the time measured by a spin clock is the same for all the considered observers, even when they move at relativistic speed. In other words, absolute time, as measured by a spin clock, is Lorentz invariant.

This result shows, in particular, that the relativistic mass effect has no impact on spin clock's time. The same conclusion can be drawn directly from Eq. (24) by noting that, as apparent from Eqs. (12), (21) and (22), the quantities $K^{spin}$ and $L_{C\ x}^{spin}$ are both proportional to the rotor mass. Their ratio, therefore, is not affected by any relativistic phenomenon that might alter the mass of the rotor itself.

Of course, for sufficiently large values of $\dot{\theta}$, the circumferential velocity $\bar{v} = r\dot{\theta}$ of the peripheral points of the rotor at a given distance *r* from its axis may be quite large, and this means that *I* may depend on $\dot{\theta}$ in a relativistic way. Reference [11] shows how to calculate *I* in this case. For large enough $\dot{\theta}$, relativistic effects may cause the value of *I* to differ from the value given by Eq. (12). The above analysis, however, proves the independence of $\dot{\theta}$ from $v$, regardless of the values of $\dot{\theta}$ or $I$, as long as these values are constant; hence, any relativistic effect on $I$ is irrelevant as long as $I$ is constant. In any case, there are no practical problems with keeping the working value of $\dot{\theta}$ of a spin clock low enough to make any relativistic effect on $I$ negligible.

Lorentz transformation applies, in particular, when the frame $[x, y, z]$ is at absolute rest (see next section) and $t$ is the absolute time. In this case, Eq. (20) expresses the relationship between the Lorentz time $t'$ and the absolute time. Lorentz time can therefore be interpreted as that particular time measure, dependent on $v$ and $x$, which makes the light velocity the same in all inertial frames. Historically, recourse to the Lorentz time may have been favoured by a lack of proof of the existence of absolute time.

**5. Absolute space**

To determine the distance between two points that are in relative motion with respect to each other or to the observer, it is necessary to know their positions at the same instant of time. If we refer to the Lorentz time $t'$, as defined in Eq. (20), we obtain different times for the same event depending on the observer's velocity and location. The simultaneity of the Lorentz time depends on the speed and position of the observer, and is therefore relative. The same is true for the Lorentz distance between two points, which becomes observer-dependent because such is the time simultaneity of their positions.

In contrast, referring to absolute time allows us to define the absolute simultaneity of any two events. The positions that two points occupy at the same absolute time can therefore be determined, and hence the absolute distance between them.

The existence of absolute space follows from the fact that from a single reference point, we can always measure the absolute distance of any other point in space, thus endowing the space with an observer-independent position for all of its points. As described below, it is not difficult to come to the same conclusion more formally, using arguments similar to those traditionally used to justify Eqs. (17) – (20).

Let $(x, y, z, t)$ and $(x', y', z', t')$ be the space-time coordinates of an event measured in two different inertial reference frames by observers $S$ and $S'$, respectively. Space and time are assumed to be homogeneous, meaning that a translation in space or time does not change the way in which an event occurs. In other words, the laws of physics are assumed to be independent of the origin we choose for the coordinates of space and time. As is well known, this assumption implies that the relationship between the coordinates $x, y, z, t$ and $x', y', z', t'$ must be linear (cf., e.g., [7], [12] and [13]). The most general form of this relationship must therefore be:

$$\begin{aligned} x' &= a_{11} x + a_{12} y + a_{13} z + a_{14} t \\ y' &= a_{21} x + a_{22} y + a_{23} z + a_{24} t \\ z' &= a_{31} x + a_{32} y + a_{33} z + a_{34} t \\ t' &= a_{41} x + a_{42} y + a_{43} z + a_{44} t \end{aligned} \qquad (25)$$

No constant term appears in these equations because we have assumed, without loss of generality, that the origins of the primed and unprimed coordinates coincide at $t = 0$.

If we refer to absolute time, then

$$t' = t. \qquad (26)$$

Hence, from Eq. (25)$_4$ we obtain

$$a_{41} = a_{42} = a_{43} = 0 \quad and \quad a_{44} = 1. \qquad (27)$$

Space is considered to be isotropic, meaning that the properties of space are the same in all directions. No restriction on the present analysis is therefore introduced if we take the axes $x'$, $y'$ and $z'$ as being parallel to the axes $x$, $y$, $z$, respectively. In this case, the primed axes coincide with the homologous unprimed axes for $t = 0$, and remain coincident if they are at rest with respect to them. This implies the restriction

$$a_{11} = a_{22} = a_{33} = 1. \qquad (28)$$

Now suppose that the reference system $x'$, $y'$, $z'$ moves with respect to the reference system $x$, $y$, $z$ at constant velocity $v_x$ in the direction of the $x$-axis. The observer $S$ sees the $x$-coordinate of the origin of reference $x'$, $y'$, $z'$ change according to the relations $x = v_x t$. To be consistent with this, Eq. (25)$_1$ must be satisfied when $x' = y' = z' = y = z = 0$, and $x = v_x t$. In view of Eq. (27)$_1$, this requires that

$$a_{14} = -v_x. \qquad (29)$$

Analogous relations, i.e.,

$$a_{24} = -v_y \quad and \quad a_{34} = -v_z, \qquad (30)$$

can be obtained by applying similar arguments to the case where the reference system $x'$, $y'$, $z'$ moves with constant velocity $v_y$ and $v_z$ in the directions of the $y$- and $z$-axes, respectively.

Under the motion considered when deriving Eq. (29), the $x$-axis is always coincident with the $x'$-axis. This means that for $v_y = v_z = 0$, Eqs. (25)$_2$ and (25)$_3$ must yield, respectively, $y' = 0$ and $z' = 0$ if $y = z = 0$ (i.e., for all points of the $x$-axis). This requires that

$$a_{21} = 0 \quad and \quad a_{31} = 0. \qquad (31)$$

Similar arguments for the cases where $v_x = v_z = 0$ and $v_x = v_y = 0$ (translation in the directions of the $y$- and $z$-axes, respectively) lead to the conclusion that:

$$a_{12} = a_{32} = a_{13} = a_{23} = 0. \qquad (32)$$

This completes the determination of all the coefficients appearing in Eqs. (25).

In conclusion, reference to absolute time reduces the general transformation in Eqs. (25) to:

$$x' = x + v_x\, t$$
$$y' = y + v_y\, t$$
$$z' = z + v_z\, t$$
$$t' = t$$
(33)

making it Galilean. From these equations, we can immediately infer that the distance $d_{AB}$ between any two points A and B in space is given by

$$d_{AB}^2 = (x'_B - x'_A)^2 + (y'_B - y'_A)^2 + (z'_B - z'_A)^2 = (x_B - x_A)^2 + (y_B - y_A)^2 + (z_B - z_A)^2, \quad (34)$$

with obvious meaning of the notation adopted. The result in Eqs. (33) therefore implies that the distance between two points in space is independent of the velocity ($v_x$, $v_y$, $v_z$) of the reference frame that we choose to describe the space. Hence, the existence of absolute time implies the existence of absolute space, as inferred at the beginning of this section in more descriptive terms.

## 6. Absolute motion detection

Maxwell's equations, whose validity is beyond question here, imply that light and, more generally, electromagnetic radiation propagates in a vacuum at a constant speed. This means that the speed of light in a vacuum is independent of the initial conditions under which the light is emitted, and, in particular, of the motion of the light source. As explained below, this well-known fact of physics means that the *absolute* position of a point that emits a light pulse can be determined from anywhere in space.

A light pulse flashing from a point $P$ in a vacuum generates a spherical wavefront whose radius $\rho$ grows at the speed of light $c$. No matter how far the wavefront travels through space, its centre remains fixed on $P$. In more descriptive terms, the wavefront keeps the memory of the absolute position of the point at which it was generated. As the wavefront passes through a point $A$ in space, we can determine the position of $P$ by taking the distance $\rho_A = c\, \Delta t$ along the normal to the wavefront from $A$, i.e., along the light ray from $P$ to $A$ (see Fig. 2). Here, $\Delta t$ indicates the absolute time interval between the emission of the flash from $P$ and the arrival of the wavefront at $A$. If $\Delta t$ is unknown, the intersection of two rays relative to two different points, shown as $A$ and $B$ in Fig. 2 (not necessarily on the wavefront at the same time), can be used to determine the absolute position of the source point $P$.

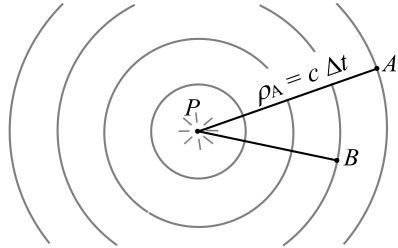

**FIG. 2.** Successive positions of the spherical wavefront generated at $P$ by a light pulse. Points $A$ and $B$ are reached at different times if they are not at the same distance from $P$.

Thus, the point *P* at which a light pulse is emitted remains fixed over time. It can be taken as a reference point at absolute rest, and its position relative to any point of space can in principle be determined as the wavefront passes through that point. The motion of the body that emits the light pulse from *P* is of no importance. The same is true for the motion of points such as *A* or *B*, which receive the wavefront. This is because the construction described above depends only on the instantaneous position of the receiving point at the instant of arrival of the wavefront. The distance $\rho_A$ determined in this way is absolute, because it is entirely obtained from measurements of absolute time, which is independent of the observer's velocity or position.

The above construction is not valid if the light emitted from *P* is continuous and uniform. In this case, the observer cannot distinguish which wavefront is emitted from *P* at a specific time (say, at *t* = 0). Furthermore, the observer cannot know *a priori* whether the point from which the uniform light comes is at absolute rest, as is always the case for the emission of a pulse of light. For instance, if there is no relative motion between the source and observer, the latter perceives the source point *P* as fixed. However, the truth is that the centre of the wavefronts reaching the observer is constantly changing over time, because *P* is not at absolute rest. In contrast, in the case of a light pulse, the centre *P* of the wavefront is always at absolute rest, regardless of the motion of the source or observer.

The absolute motion of a reference frame [*O*, *x*, *y*, *z*] can be determined by an observer inside the frame. The observer must know the absolute distance of the point *P* from which a light pulse is emitted and the absolute time of emission. Point *P* can be stationary or moving; it could be a star, a planet, or a point on a body rigidly connected to the reference frame under consideration. The absolute time at which the flash is emitted is necessary in order to measure the absolute time taken by the wavefront to reach the observation point.

More specifically, let us assume that the light is emitted from *P* at absolute time *t* = 0 and that the observer makes measurements at the origin *O* of his or her frame. To simplify, we suppose that emission of the flash occurs when the point *P* is on the *y*-axis, at a known distance *h* from *O*. As discussed above, the absolute position of the centre of the spherical wavefront from *P* remains fixed at *P*, regardless of the distance of wavefront from *P*. If the frame [*O*, *x*, *y*, *z*] is at absolute rest, the wavefront from *P* reaches point *O* at the absolute time $t_0 = h/c$, with a radius *PO* normal to the *x*-axis (Fig. 3). The observer finds that at time $t_0$ the angular height of *P* is $\varphi_0 = \pi/2$ above the *x*-axis, coinciding with the value it had at time *t* = 0. From the observed values of $t_0$ and $\varphi_0$, the observer can conclude that his or her frame is at absolute rest.

Let us suppose instead that the reference frame moves and that at *t* = 0 its origin has an absolute velocity $v_x$ in the direction of the *x*-axis. In this case, the wavefront from *P* reaches the origin of the frame at time $t_1 > t_0$, at a point $O_1$ on the *x*-axis at an absolute distance $d = v_x t_1$ from *O*. At the same time $t_1$, the absolute distance of *P* from $O_1$ is $\rho = c t_1 > h$. In Fig. 3, the position of the reference frame at time $t_1$ is represented as [$O_1$, $x_1$, $y_1$, $z_1$]. In this case, the observer finds that the value $\varphi_1$ of the angular height of *P* above the *x*-axis is different from the value $\varphi_0$ that applied at time *t* = 0 or when the frame is at absolute rest at *O*.

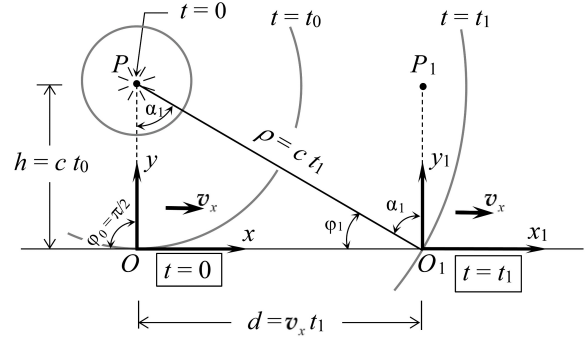

**FIG. 3.** A reference frame moving in the *x*-direction receives the wavefront of a light pulse from point *P* at different times and with different elevation angles φ, depending on the absolute velocity $v_x$ of the frame.

As shown in Fig. 3, we have $d = \rho \cos \varphi_1$. Thus, if

$$\alpha_1 = \frac{\pi}{2} - \varphi_1 \qquad (35)$$

denotes the complementary angle to $\varphi_1$, it follows from the definition of $d$ and $\rho$ that

$$\sin \alpha_1 = \frac{d}{\rho} = \frac{v_x}{c}, \qquad (36)$$

that is,

$$v_x = c \sin \alpha_1. \qquad (37)$$

Hence, by measuring $\varphi_1$, or, equivalently, $\alpha_1$, the observer can determine the absolute velocity of the frame in the direction of the *x*-axis.

A similar procedure can be followed to determine the absolute motion of the reference frame in the *y* and *z* directions. For instance, if the $v_z$ component of the velocity is sought, we could retain the same light source *P* on the *y*-axis shown in Fig. 3, but consider the *z*-axis rather than the *x*-axis considered above. To determine $v_y$, we could consider a light source on one of the other two axes, say the *z*-axis, and apply the above procedure by referring to the *y*- and *z*-axes instead of the *x*- and *y*-axes, respectively.

Since the velocity of the observer is much less than *c*, the angle $\alpha_1$ is usually very small, as is evident from Eq. (36)$_2$. This may make the practical application of the above procedure quite delicate; its conceptual importance, however, should not be underestimated. In principle, the entire procedure can be executed in a "box" that is entirely blind to the surroundings. The source point *P* may be inside the box, attached to one of its walls and therefore moving with it, and the box may be moving at a constant speed, thus embodying an inertial reference frame. Therefore, by exploiting the properties of light propagation in a vacuum, the procedure presented above gives us the means to determine the absolute velocity of the box, without making any reference to the surrounding world.

## 7. Conclusions

Absolute time, independent of the velocity and position of the observer, is a consequence of the law of conservation of angular momentum. It has been used for centuries in classical physics, and is a fundamental quantity that is suitable for measuring time in all fields of physics in which the law of angular momentum balance holds. The special theory of relativity is one of these fields, as evidenced in Section 4.

The existence of absolute time implies the possibility of establishing the absolute simultaneity of two events, and hence of measuring the absolute distance between any two points in space. As a consequence, absolute time implies the existence of absolute space.

In principle, the position of a point in absolute space can be determined by an observer by exploiting the independence of the velocity of light in a vacuum from the motion of its source. To do this, the observer must know the absolute time at which a light pulse is emitted from a given point, and must also determine the absolute time of the arrival at the observation point of the wavefront produced by the light pulse and the direction of the incoming ray.

On the other hand, the state of absolute motion or rest of a reference frame can, at least in theory, be detected by observing the wavefront emitted in a vacuum by a pulse of light produced by a point moving with the frame. The observer need not refer to points outside the frame, although the position of the observer relative to the source of light at the moment of the flash and the direction of the light ray as it arrives at the observation point must be known.

Finally, it may be worth stressing that time, as discussed in this paper, is defined with reference to a physical system—a clock—that performs a stable cyclic process. The paper demonstrates that it is possible to construct a clock that, even at relativistic speeds, provides the same measure of time regardless of its motion or location. This clock measures absolute time, as currently understood in physics. The existence of such a clock proves the existence of absolute time and space.